\documentstyle[12pt,twoside,fleqn,espcrc1,epsfig,rotate]{article}
\title{Excitation function of strangeness in A+A reactions from SIS to RHIC
energies\footnote{Supported by BMBF and GSI Darmstadt}}
\author{W. Cassing\thanks{In collaboration with E. L. Bratkovskaya, J. Geiss,
C. Greiner, S. Juchem and U. Mosel}\\
Institut f\"ur Theoretische Physik, Universit\"at Giessen \\
D-35392 Giessen, Germany}

\begin{document}
\maketitle

\begin{abstract}
The properties of $K^+$ and $K^-$ mesons are studied in nuclear
reactions from SIS to RHIC energies within the covariant transport
approach HSD in comparison to the experimental data whenever
available. Whereas kaon abundancies and spectra indicate little
repulsive or vanishing selfenergies in the medium, antikaons are
found to experience strong attractive potentials in
nucleus-nucleus collisions at SIS energies. However, even when
including these potentials the $K^+$ and $K^-$ spectra at AGS
energies are noticeably underestimated showing an experimental
excess of strangeness that points towards a nonhadronic phase in
these reactions.  On the other hand the $K^+, K^-$ production at
SPS energies is again well described by the HSD approach
based on quark, diquark, string and hadronic degress of freedom. At RHIC
energies hadronic rescatterings are found to enhance the strangeness yield from
the partonic phase quite substantially.
\end{abstract}

\section{INTRODUCTION}
The aim of high energy heavy-ion collisions is to investigate
nuclear matter under extreme conditions, i.e. high temperature and
density. The most exciting prospect is the possible observation of
a signal for a phase transition from normal nuclear matter to a
nonhadronic phase, where partons are the basic degrees of freedom.
In this context strangeness enhancement in heavy-ion collisions
compared to proton-proton collisions has been suggested as a
possible signature for the phase transition \cite{Rafelski1}. On
the other hand, precursor effects might already be seen at SIS
energies since densities up to 3$\times \rho_0$ can be achieved in
central collisions of heavy nuclei  and the effect of meson
potentials can be studied with a higher sensitivity to the
productions thresholds, respectively. In this contribution a brief
survey is presented on the information gained so far in comparison
of experimental data to nonequilibrium transport theory, here the
Hadron-String-Dynamics (HSD) approach \cite{CBPR98}. For a more
detailed discussion of the issues presented the reader is refered
to \cite{CBPR98,Geiss}.

\section{ANALYSIS OF EXPERIMENTAL DATA}
Since the real part of the actual $K^+$ and $K^-$ self-energy
$\Pi_{K^\pm}$ in hot and dense nuclear matter is quite a matter of
debate we adopt a more practical point of view and as a guide for
the analysis use a linear extrapolation of the kaon potential with
density $\rho_B$ as
\begin{equation}
\label{kmass} m^*_K(\rho_B) = m_K^0 \left(1 - \alpha
\frac{\rho_B}{\rho_0}\right), \ {\rm i.e.} \ V_K = - \alpha m_K^0
\frac{\rho_B}{\rho_0},
\end{equation}
with $\alpha_{\bar{K}} \approx $ 0.2-0.25 for antikaons and
$\alpha_K \approx -0.06$ for kaons in line with Refs. \cite{Schaffner,waas}.
In (\ref{kmass}) a momentum
dependence of the kaon or antikaon potential has been neglected
for reasons of numerical simplicity. The dispersion analysis of
Sibirtsev et al. \cite{Sibdr} shows that this is roughly fulfilled
for the kaon potential, however, the antikaon potential should be
more strongly momentum dependent.

\begin{figure}[t]
\phantom{a}\vspace*{-10mm}
\centerline{\psfig{figure=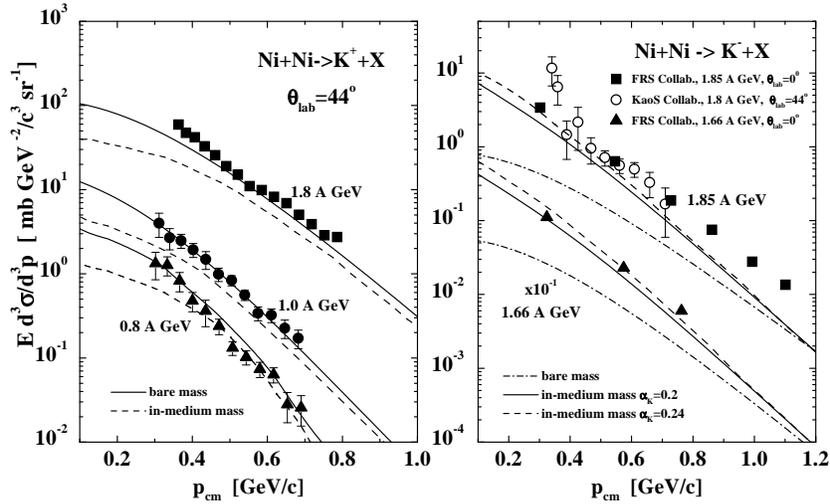,width=11.5cm} }\vspace*{-15mm}
\caption{ The calculated $K^+$ (l.h.s.) and $K^-$ (r.h.s.)
momentum spectra in the nucleus-nucleus cms for Ni~+~Ni reactions
for different meson potentials in comparison to the experimental
data (see text).} \label{Fig2}
\end{figure}

 The Lorentz invariant $K^+$ spectra for
Ni~+~Ni at 0.8, 1.0 and 1.8~A$\cdot$GeV are shown in
Fig.~\ref{Fig2} (l.h.s.) in comparison to the data from the KaoS
Collaboration \cite{Senger}.  Here the full lines reflect
calculations including only bare $K^+$ masses ($\alpha_K = 0$)
while the dashed lines correspond to calculations with $\alpha_K =
- 0.06$ in Eq.~(\ref{kmass}), which leads to  an increase of the
kaon mass at $\rho_0$ by about 30~MeV.  The general tendency seen
at all bombarding energies is that the calculations with a bare
kaon mass seem to provide a better description of the experimental
data for Ni~+~Ni than those with an enhanced kaon mass. This trend
continues to hold also for the light system C~+~C as well as for
the heavy systems Ru + Ru \cite{FOPI} and even Au~+~Au
\cite{CBPR98} within the cross sections for the $\Delta$ induced
channels used.

The kaon flow in the reaction plane shows some sensitivity to the
kaon potential in the nuclear medium as suggested by Li, Ko and
Brown \cite{flow1}.  Here due to elastic scattering with nucleons
the kaons partly flow in the direction of the nucleons thus
showing a positive flow in case of no mean-field potentials.
  With increasing repulsive kaon potential the
positive flow will turn to zero and then become negative. In fact,
experimental data on kaon flow indicate a repulsive
potential for kaons in the nuclear medium at SIS \cite{Brat622} as
well as AGS energies \cite{Rai}.

We now turn to the production of antikaons which do clearly show
the effect from attractive potentials in the medium. We recall
that for $\alpha_{\bar{K}}$ = 0 in Eq. (\ref{kmass}) we recover
the limit of vanishing antikaon self-energy, whereas for
$\alpha_{\bar{K}} \approx$ 0.2  we approximately describe the
scenario of Refs. \cite{Schaffner,waas}.  The $K^-$ spectra for Ni~+~Ni
at 1.85 and 1.66 A$\cdot$GeV \cite{Schro}
are shown in Fig. \ref{Fig2} (r.h.s.)
for $\alpha_{\bar{K}}$ = 0, 0.2 and 0.24 where the latter cases
correspond to an attractive potential of $-100$ and $-120$~MeV at
density $\rho_0$, respectively.  We note, that due to the
uncertainties involved in the elementary $BB$ production cross
sections we cannot determine this value very reliably.  With
increasing $\alpha_{\bar{K}}$ not only the magnitude of the
spectra is increased, but also the slope becomes softer. This is
most clearly seen at low antikaon momenta because the net
attraction leads to a squeezing of the spectrum to low momenta \cite{FOPI}.

 Whereas the kaon and antikaon dynamics at SIS energies is
reasonably described within the hadron-string transport approach
when including meson potentials \cite{CBPR98}, this no longer
holds at AGS energies \cite{Geiss}. The heaviest system studied
here is Au~+~Au at $\approx$ 11 A$\cdot$GeV. In the 'bare mass'
scenario the data are underestimated strongly while for $\alpha_K
= -0.06$ and $\alpha_{\bar{K}}$ = 0.24 the situation improves
significantly \cite{CBPR98}. Whereas the $K^-$ yield is almost
reproduced in the latter scheme, the $K^+$ yield is still
underestimated as in case of the Si~+~Al and Si~+~Au system at
14.6 A$\cdot$GeV \cite{CBPR98,Geiss}.

Without explicit representation we note that for all systems at
SPS energies \cite{Geiss} the $h^-$, $K^+$ and $K^-$ distributions
are reproduced rather well showing even a tendency for an excess
of kaons and antikaons in the calculations rather than missing
strangeness. At RHIC energies we also find a sizeable enhancement
of the $K^{\pm}$ yield in central Au~+~Au compared to $pp$
collisions or pure parton (VNI) cascade calculations due to a long
lasting hadronic rescattering phase \cite{CBPR98}.

\begin{figure}[t]
\phantom{a}\vspace*{-5mm}
\centerline{\psfig{figure=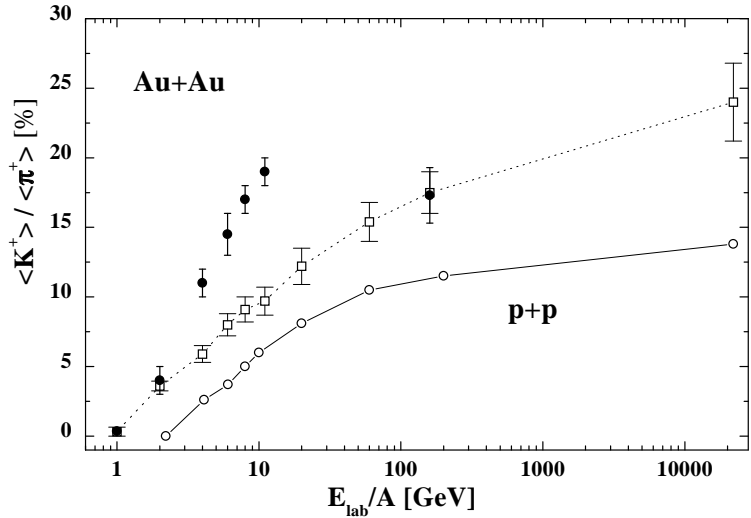,width=11cm}}
\vspace*{-10mm} \caption{The calculated $K^+/\pi^+$ ratio at
midrapidity for central Au~+~Au reactions (open squares) from SIS
to RHIC energies in comparison to the preliminary experimental
data from 1 - 160 A$\cdot$GeV and the corresponding ratio for p +
p collisions (open circles) from the HSD approach (see text). }
\label{Fig5}
\end{figure}

The E866 and E895 Collaborations, furthermore, have measured
Au~+~Au collisions at 2,4,6 and 8 A$\cdot$GeV kinetic energy at
the AGS \cite{Ogilvi}.  Thus it is of particular interest to look
for a {\em discontinuity in the  excitation functions} for pion
and kaon rapidity distributions and to compare them to the
hadron-string transport approach.  In Fig. \ref{Fig5} the
calculated $K^+/\pi^+$ ratios (open squares) at midrapidity
($|y_{cm}| \leq$ 0.25) for central (b=2 fm) Au~+~Au collisions at
1,2,4,6,8 and 11 A$\cdot$GeV and Pb~+~Pb collisions at 160 A$\cdot$GeV
are shown together with the preliminary data (full
dots). The ratio at midrapidity is slightly higher than the total
$K^+/\pi^+$ ratio, because the kaon rapidity distribution is
narrower than that of the pions.  While the scaled kaon yield at 1
and 2 A$\cdot$GeV (SIS energies) is well described in the HSD
approach within the errorbars, the experimental $K^+/\pi^+$ ratio
at 4 A$\cdot$GeV is underestimated already by a factor of 2 and
increases up to roughly 19\% for 11 A$\cdot$GeV.  As mentioned
before the calculated and measured ratio coincide again at 160
A$\cdot$GeV. Data at 40 A GeV from the SPS as well as at 21.5 A
TeV from RHIC are expected to come up soon and to complete the
picture from the experimental side.

\section{SUMMARY}
We find an enhancement of the $K^+/\pi^+$ ratio in heavy-ion
collisions relative to p~+~p reactions  due to hadronic
rescatterings both with increasing system size and energy.
 The excitation function in the $K^+/\pi^+$ ratio from the
HSD transport approach has a similar slope in nucleus-nucleus and
p~+~p collisions (cf. Fig. 2) indicating a monotonic increase of
strangeness production with bombarding energy. However, the
experimental $K^+/\pi^+$ ratio for central Au~+~Au collisions at
midrapidity increases up to $\approx 19\%$ at 11 A$\cdot$GeV -- it
is unknown if a local maximum will be reached at this energy --
and decreases at SPS energies to $\approx 16.5\%$.  Such a
decrease of the scaled kaon yield from AGS to SPS energies is hard
to obtain in a hadron-string transport model. On the contrary, the
higher temperatures and particle densities at SPS energies allways
tend to enhance the $K^+/\pi^+$ yield closer to its thermal
equlibrium value of $\approx 20-25\%$  at chemical freezout and
temperatures of $T\approx 160$ MeV.  Thus the steep rise of the
strangeness yield and its decrease suggests the presence of
nonhadronic degrees of freedom which might become important
already at about 4 A$\cdot$GeV in central Au~+~Au collisions.

\end{document}